\documentclass[lettersize,journal]{IEEEtran}
\usepackage{amsmath,amsfonts}
\usepackage{algorithmic}
\usepackage{algorithm}
\usepackage{array}
\usepackage[caption=false,font=normalsize,labelfont=sf,textfont=sf]{subfig}
\usepackage{textcomp}
\usepackage{stfloats}
\usepackage{url}
\usepackage{verbatim}
\usepackage{graphicx}
\usepackage{cite}
\usepackage{hyperref}
\usepackage{booktabs}
\usepackage{subfig}

\begin{document}

\title{Empathic Responding for Digital Interpersonal Emotion Regulation via Content Recommendation}

\author{\IEEEauthorblockN{Akriti Verma\IEEEauthorrefmark{1},
Shama Islam\IEEEauthorrefmark{1},
Valeh Moghaddam\IEEEauthorrefmark{2},
Adnan Anwar\IEEEauthorrefmark{2},
Sharon Horwood\IEEEauthorrefmark{3},}

\IEEEauthorblockA{\IEEEauthorrefmark{1}School of Engineering, Deakin University, Australia}
\IEEEauthorblockA{\IEEEauthorrefmark{2}School of Information Technology, Deakin University, Australia}
\IEEEauthorblockA{\IEEEauthorrefmark{3}School of Psychology, Deakin University, Australia}}




\maketitle

\begin{abstract}
Interpersonal communication plays a key role in managing people's emotions, especially on digital platforms. Studies have shown that people use social media and consume online content to regulate their emotions and find support for rest and recovery. However, these platforms are not designed for emotion regulation, which limits their effectiveness in this regard. To address this issue, we propose an approach to enhance Interpersonal Emotion Regulation (IER) on online platforms through content recommendation. The objective is to empower users to regulate their emotions while actively or passively engaging in online platforms by crafting media content that aligns with IER strategies, particularly empathic responding. The proposed recommendation system is expected to blend system-initiated and user-initiated emotion regulation, paving the way for real-time IER practices on digital media platforms. To assess the efficacy of this approach, a mixed-method research design is used, including the analysis of text-based social media data and a user survey. Digital applications has served as facilitators in this process, given the widespread recognition of digital media applications for Digital Emotion Regulation (DER). The study collects 37.5K instances of user posts and interactions on Reddit over a year to design a Contextual Multi-Armed Bandits (CMAB) based recommendation system using features from user activity and preferences. The experimentation shows that the empathic recommendations generated by the proposed recommendation system are preferred by users over widely accepted ER strategies such as distraction and avoidance.

\end{abstract}



\section{Introduction}
Today, social media has seamlessly integrated into our daily lives, and individuals leverage digital technologies, including social media platforms, instant messaging, and online communities, to navigate their emotions within online social interactions and relationships \cite{smith2022digital}, \cite{prikhidko2020effect}. Platforms like Facebook, Twitter, Reddit, Instagram and online support communities serve as avenues for sharing emotions and seeking emotional support, thereby acting as sources of emotional regulation within digital spaces \cite{costello2017confronting}. The interplay of activity, feedback, and responses from online social networks becomes vital in shaping emotional well-being, necessitating an examination of its potential to either exacerbate or alleviate emotional distress \cite{kubin2021role}, \cite{kramer2014experimental}.

As the prevalence of digital interpersonal emotion regulation grows, the need to understand its impact on emotional well-being becomes increasingly evident. Several studies on social media activity and user emotions have revealed the complex relationship between political discourse, emotion regulation, and the expression of political and other emotions on social media \cite{kubin2021role}, \cite{duncombe2019politics}. A focus on charismatic leaders and their communication strategies on platforms like Twitter offers a unique perspective on how emotion regulation is employed as a tool for audience engagement. Furthermore, it sheds light on the role of empathy in the dissemination and regulation of these emotions \cite{henry2023express}, \cite{kissas2020performative}.

Recent studies in psychological literature emphasise the value of empathic responding in managing emotions and fostering positive interactions on social media. Field experiments, such as the one conducted by Hangartner et al., reveal the potential of empathy-based counter speech to mitigate hate speech and promote constructive discussions \cite{hangartner2021empathy}. Similar findings by Nozaki and Mikolajczak et al. highlight the efficacy of reappraisal and empathic responding in alleviating others' negative emotions during text-based online interactions \cite{nozaki2022effectiveness}.

Prior work in the domain of digital interpersonal emotion regulation has focused on crafting tools for assisting users in providing support to others through empathy and appreciation \cite{pradana2017imparting}. More recently, there has been a shift towards acknowledging and addressing user emotions directly, through recommendation systems like emotion-based music and movie recommendation systems \cite{james2019emotion}, \cite{ayata2018emotion}. These systems employ various sources of emotional data, including physiological sensors, facial expressions, and fuzzy emotion features, to curate personalised recommendations for the user's emotional state.

However, it's crucial to distinguish these systems from those designed for emotion regulation, as their primary focus lies in satisfying the intrinsic emotional needs of the user, emphasising personalised emotional alignment rather than active emotion regulation. In this context, Ameko et al. introduced a novel treatment recommendation system dedicated to emotion regulation, leveraging real-world historical mobile digital data through contextual bandits \cite{ameko2020offline}, \cite{beltzer2022building}. A contextual multi-armed bandit (CMAB) algorithm works by integrating contextual or supplementary information for decision-making contexts to effectively manage the trade-off between exploration (experimenting with new options) and exploitation (leveraging known information to make optimal choices). They incorporated physiological signals and interactive question-based data gathered from individuals experiencing anxiety to generate tailored recommendations for emotion regulation strategies, for people experiencing anxiety issues. However, emotion regulation can be undertaken while simply scrolling through a digital media application \cite{hossain2022motivational}, \cite{sarsenbayeva2020does}.

Therefore, this paper proposes a novel recommendation system that integrates content preferences with emotion regulation strategies (hence initiating ER). Acknowledging the challenge posed by limited contextual cues in social media platforms, the model refrains from prescribing specific strategies. Instead, it empowers users by providing content recommendations aligned with emotion regulation techniques, particularly empathic responding. This approach not only curtails mindless scrolling and mitigates emotional distress but also enhances the user's awareness of the influence that online content has on their emotional states. By merging content preferences with emotion regulation strategies, the recommendation system aims to make the process of emotion regulation more intuitive, ultimately empowering users to actively and effectively manage their emotions in the digital landscape \cite{zsido2021role} .

We conducted a mixed-method study utilising text-based social media data to generate customised empathetic content and then validate its efficacy via a survey-based user study. This research aims to contribute to a deeper understanding of the dynamics of individual emotional profiles in online settings, their effects on social media activity, and the potential for cultivating positive emotional experiences within digital communities. Therefore, the main contributions of this work are: 
\begin{itemize} 
    \item We present a framework for digital interpersonal emotion regulation via content recommendation, designed to facilitate effective emotion regulation while scrolling online. This model incorporates empathic responding to enhance user experience and emotional well-being.
    \item We introduce a recommendation system that leverages user content preferences and emotional alignment. This system supports active regulation of user emotions by recommending content based on empathic responding, demonstrated through the analysis of 37.5K user posts and interactions from Reddit.
    \item We conducted a user study to assess how users responded to the empathetic content generated by the proposed recommendation system. Our findings indicated that users preferred empathic recommendations over other strategies like distraction and avoidance.
\end{itemize}

\section{Literature Review}
This section talks about the current research and methodologies in the field of emotion regulation within digital environments. It focuses on three key areas: the mechanisms and strategies for emotion regulation online and the development and impact of emotion-based content recommendation systems.

\subsection{Emotion Regulation Online}
Emotion regulation is a critical aspect of human psychology, and plays a critical role in mental health and interpersonal relationships \cite{gross2008emotion}, \cite{gross1998emerging}. The advent of digital media platforms has sparked significant interest in exploring how these platforms can facilitate emotion regulation processes. Digital media provides opportunities for individuals to seek and receive emotional support, share experiences, and access resources that can help in emotion regulation \cite{baker2016relationship}. Recent research has investigated how individuals utilise various social media applications to regulate their emotions, thoughts, and behaviours. These studies have explored various aspects of everyday emotion regulation, including the use of social media to alleviate homesickness and the role of music streaming platforms among university students \cite{smith2022digital}, \cite{sarsenbayeva2020does}, \cite{wadley2019use}. They shed light on multitasking and passive scrolling habits, and how social media breaks impact emotional well-being. Researchers have also compared active versus passive social media usage \cite{hossain2022motivational}. Recent interventions have utilised biofeedback and haptic interactions to offer real-time and offline support. Some apps provide on-the-spot guidance with triggers and create customised emotional regulation journeys by visualising emotional timelines \cite{smith2022digital}. There are also didactic interventions that rely on reminder-based recommender systems to suggest strategies and encourage reflection. These platforms shape Digital Emotion Regulation, improving well-being through technology and open up new research areas. 

Interpersonal emotion regulation refers to the processes through which individuals regulate their own or others' emotions within interpersonal contexts \cite{niven2009classification}. The role of interpersonal interactions in emotion regulation has been well-documented, highlighting the importance of social support and empathic communication in facilitating emotional well-being \cite{zaki2013interpersonal}. Digital media platforms offer an arena for interpersonal emotion regulation, enabling individuals to engage in empathic interactions and receive emotional support from a broader network of peers \cite{reis2000relationship}, \cite{costello2017confronting}. Users interact with each other through posting, commenting, reacting, or messaging online, all of which are avenues for interpersonal emotion regulation \cite{kubin2021role}, \cite{kramer2014experimental}. Uninformed interpersonal emotion regulation is a common occurrence on social media platforms. This phenomenon is often used for intentional mass emotion regulation, which is known as emotional contagion. However, uninformed or unintentional emotion regulation can lead to maladaptive emotion regulation, and users who passively engage on social media are particularly susceptible to this \cite{henry2023express}, \cite{kissas2020performative}. Furthermore, it may not always be possible for a person to be available to provide emotional support. Hence, we propose a system that generates empathic content based on user preferences to help users regulate their emotions.

\subsection{Emotion based Content Recommendation}
Empathy, the ability to understand and share others' emotions, is a vital component of effective communication and emotional regulation \cite{decety2014complex}, \cite{preece1999empathy}. Incorporating empathic content into digital platforms has shown positive results in boosting emotional well-being and fostering connections between individuals \cite{moreno2011feeling}, \cite{tejaswinimusic}, \cite{james2019emotion}, \cite{ayata2018emotion}, \cite{rumiantcev2020emotion}. Empathic content can create an environment where users feel supported and acknowledged, facilitating effective emotional regulation strategies \cite{rime2009emotion}.
Individuals utilise different emotional regulation (ER) strategies in response to stressful events based on varying social and physical situations and situational demands. Recent research has shown that the effectiveness of emotion regulation (ER) strategies, previously categorised as adaptive or maladaptive, is heavily influenced by the context in which they are employed \cite{zsido2021role}. Studies have revealed that internal contextual factors, such as age and gender, significantly impact individual decision-making concerning ER strategies. Additionally, external contextual factors have also been investigated, and it has been found that common ER strategies such as cognitive reappraisal may not always be the most effective in mitigating negative emotions \cite{ameko2020offline}, \cite{heiy2014back}. Therefore, it is imperative to identify and recommend effective ER strategies to individuals based on their particular contextual factors. In this work, we aim to achieve this objective by developing a personalised and adaptable approach for recommending content based on ER strategies.

\begin{table*}[htbp]
\centering
\caption{Comparison of Emotion-Based Recommendation Systems}
\label{tab:emotion_recommendation}
\begin{tabular}{p{2.5cm}p{2.5cm}p{2.5cm}p{2cm}p{3cm}p{3cm}}
\toprule
Reference & Data or user attributes & Recommendation Approach & Target Content & Key Features & Limitations \\
\midrule
Emotion based music recommendation system \cite{lahariemotion}, \cite{agrawalmoodplayer}, \cite{quah2024music}, \cite{swathi2024research}, \cite{tejaswinimusic}, \cite{james2019emotion}, \cite{ayata2018emotion}, \cite{rumiantcev2020emotion} & Facial expressions, physiological signals, user surveys, speech emotion recognition & Linear mapping, Deep neural networks, LSTM, CNN, clustering, collaborative or content-based recommendation, SVM & Music & Improved user satisfaction with recommended content, enhanced accuracy, situation-based music selection, enhanced playlist creation & Focus on enhancing mood, providing support for a single strategy of regulating emotions \\
Emotion based movie recommendation system \cite{saraswat2020analyzing}, \cite{ho2006mrs} & Fuzzy emotion features from text reviews & Collaborative or content-based recommendation & Movies & Improved user satisfaction with recommended content & Focus on enhancing mood \\
Emotion based content recommendation system \cite{babanne2020emotion} & Facial feature extraction & Content-based and collaborative filtering & Video content & Improved user satisfaction with recommended content and enhanced prediction accuracy & Focus on enhancing mood and prediction accuracy \\
Knowledge and sentiment-based recommendation system \cite{rosa2018knowledge} & Emotions from text and user profile information and location & Sentiment analysis, deep learning & Text messages & Detects depressive or stressed content & Designed to provide warning, information, or flag depressive content \\
Treatment recommender system for emotion regulation \cite{ameko2020offline} & EMA and historical mobile digital data & CMAB & ER strategies & Helps with anxiety by recommending strategies for ER & Designed for a specific use case of anxiety experiencing individuals \\
\textbf{Empathic Responding for Digital Interpersonal Emotion Regulation via Content Recommendation} & \textbf{User activity, personality, emotions features in text, current emotional state} & \textbf{CMAB} & \textbf{Text-based content} & \textbf{Initiates interpersonal emotion regulation by recommending empathic content based on user features and current emotion} & \textbf{Needs to be tested on a broader audience.} \\
\bottomrule
\end{tabular}
\end{table*}

Recommendation systems are integral to shaping user experiences on digital platforms, offering personalised content based on user preferences and behaviours \cite{schafer1999recommender}. Contextual recommendation systems, in particular, have gained popularity due to their ability to incorporate additional information such as user emotions, preferences, and situational factors, leading to more personalised and empathic content recommendations \cite{adomavicius2010context}, \cite{raza2019progress}. These systems have significant potential to promote interpersonal emotional regulation by providing content that aligns with users' emotional needs and regulatory strategies.

As shown in Table-\ref{tab:emotion_recommendation}, recent literature has seen the development of several emotion-based recommendation systems, aimed at improving user satisfaction and accuracy in content recommendations across different domains. These systems use a variety of methods to detect emotions, including facial expressions, physiological signals, sentiment analysis, and deep learning techniques. However, most existing systems focus on enhancing mood through content recommendations, neglecting the broader range of emotion regulation strategies available. In contrast, the proposed recommendation system, `Empathic Responding for Digital Interpersonal Emotion Regulation via Content Recommendation,' integrates user activity, personality traits, and current emotional state to trigger system-initiated interpersonal emotion regulation. By suggesting empathic content tailored to individual features and emotions, this system aims to facilitate effective emotion regulation strategies beyond mood enhancement, providing an approach for emotional well-being in digital spaces. While music is a widely used tool for regulating emotions, it is important to note that it offers only one way of performing emotion regulation, namely expression. Although expression is a favoured strategy for regulating a few emotions, it may not be effective for all, as shown later in our experiment.

The field of emotion-based recommendation systems has made significant progress, but there are still notable gaps in addressing the regulation of emotions. 
\begin{itemize}
    \item One key area that lacks attention is emotion regulation. Current recommendation systems primarily aim to match content with users' current emotions, rather than actively helping users regulate their emotions. Research suggests that there is a need for systems that can support users in moving from one emotional state to another more desired state, as users often turn to digital media for emotion regulation \cite{wadley2020digital},  \cite{slovak2023designing}.
    \item Furthermore, the role of empathic responding in digital environments has not been thoroughly explored. There is a gap in developing systems that can provide empathic responses and support through content recommendations, thereby enhancing interpersonal emotion regulation.
\end{itemize}

To address these gaps, the current study focuses on developing a content recommendation system that targets the regulation of emotions using Contextual Multi-armed Bandits (CMABs). Unlike existing systems that align content with users' current emotions, the proposed system leverages CMABs to recommend content that helps users regulate their emotions through strategies such as empathic responding, distraction, avoidance, expression, and relaxation.

\section{Proposed Framework}
\begin{figure}[h]
  
    \centering
    \includegraphics[width=8.5cm,height=14cm,keepaspectratio]{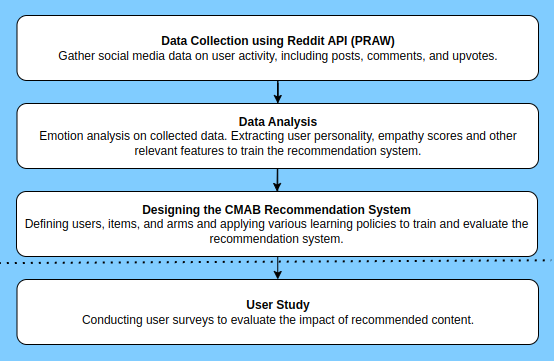}
  \caption{Proposed high-level framework for empathic responding via content recommendations}
  \label{fig:method}
  \end{figure}

The proposed framework, as depicted in Fig. \ref{fig:method}, designs empathic recommendations on social media by utilising two key components. The first component entails a quantitative analysis of Reddit user behaviour through the \href{https://praw.readthedocs.io/en/stable/}{Reddit API (PRAW)} to gather and explore social media data (detailed in part B of this section). This data centres on user activity, including posts, comments, and upvotes. The acquired data was then used for emotion analysis, extraction of user personality and empathy scores, and other features that would be employed to train the recommendation system.

The second element of this research consists of a user study that utilises surveys to evaluate the impact of recommended content on social media platforms (detailed in part E of this section)\cite{tag2021retrospective}. Drawing on research on extrinsic emotion regulation, the study aims to determine the effectiveness of empathetic responses in mitigating negative emotions experienced by users \cite{nozaki2022effectiveness}. Additionally, it investigates whether the effectiveness of this approach is influenced by the intensity of the emotions expressed. Participants were asked to share their feedback through a Qualtrics-based survey, which included questions about the recommended content and validated scales that assess the efficacy of social media content for empathetic responses (refer to Table \ref{tab:survey_questions} which presents the survey questions for the user study). Following this, the effectiveness of the generated empathetic recommendations was analysed.


We will use the following terms in this paper: (i) action/arm: recommendation candidates (posts), (ii) reward: customer interaction from a single trial, such as a click or upvote, (iii) value: the estimated long-term reward of an arm over multiple trials, and (iv) policy: the algorithm/agent that chooses actions based on learned values.

Building a recommender using CMAB, requires data about a set of users U, a set of items I, and their interactions R. The data used is described below.

\subsection{Data Collection and Preprocessing}

For training and evaluating the recommender algorithm, we relied on a combination of structured tabular and unstructured text data obtained from Reddit, a social news aggregation, content rating, and forum network. Specifically, we focused on text-based posts within the platform and collected a dataset comprising 375,350 rows of user activity and statistics on the platform.

The process of collecting data begins with gathering text-based social media data. Reddit is a highly popular platform where users engage in in-depth discussions, offer support, and share their views on current events \cite{manikonda2018twitter}. Our approach involves obtaining a diverse dataset that includes text-based interactions from Reddit, which we acquired using the \href{https://praw.readthedocs.io/en/stable/}{Reddit API (PRAW)}. Our goal is to capture a wide range of emotions expressed in various contexts by gathering user details and fetching their recent interactions and activity on the platform. We only consider posts containing text and emoticons, and the analysis focuses solely on English-language content while excluding all other languages. CSV files were used to save the data, with one file per user. Afterwards, the data was cleaned to remove extraneous characters, links, and special symbols, and tokenize it for easier analysis \cite{bird2009natural}.

Posts on Reddit are organised by subject into user-created boards called ``communities" or ``subreddits". Reddit provides its users with a variety of features to engage with content and other users. Posts on Reddit can be upvoted and downvoted, commented upon, awarded, and shared. Users earn karma, a score that reflects their contributions to the Reddit community, through actions such as receiving upvotes, giving or receiving awards, and contributing positively to the platform. Conversely, receiving downvotes leads to a decrease in karma. 

\subsubsection{User}
The first set of user data extracted from Reddit comprised of user details such as the user ID, subreddits a user is active in, karma and the list of posts and comments the user has made on the platform. Due to the Reddit API quota restrictions, we restricted to the 100 recent posts by a user. Table. \ref{tab:user_features} describes the final list of features in the extracted user data. 
We used this data to further extract more features for the users.
\begin{table*}[htbp]
\centering
\caption{User Features Used in the Recommendation System}
\label{tab:user_features}
\begin{tabular}{lp{10cm}}
\toprule
\textbf{Feature}                  & \textbf{Description}                                                                                                              \\
\midrule
User ID                           & Unique identifier assigned to each user.                                                                                          \\
Subreddits                       & List of subreddits where the user is active.                                                                                      \\
Karma                             & Reddit's scoring system indicating the user's contribution and engagement level on the platform.                                 \\
Recent Posts                     & The 100 most recent posts made by the user.                                                                                       \\
Recent Comments                  & The 100 most recent comments made by the user.                                                                                    \\
Emotion and Sentiment            & Analysis of the emotional tone and sentiment expressed in user comments based on the NRC Word-Emotion Association Lexicon.     \\
Big Five Personality Types       & Classification of users into personality types based on the emotional content of their comments.                                 \\
Empathy Score                    & Vector representation of user empathy generated using word2vec model, capturing the empathetic nature of the user's language. \\
Comment Features                 & Features extracted from user comments providing insights into prevalent topics and themes in user discussions.                   \\
Aggregated Comment Vectors       & Aggregated vector representations of user comments across all posts, summarizing the content and sentiment of their interactions. \\
Subreddit Interactions           & Indicator of user interactions with posts from specific subreddits associated with emotion regulation strategies.                 \\
\bottomrule
\end{tabular}
\end{table*}

To infer users' personality types, we use the emotion and sentiment that match a particular word from the user's comments, using the dictionary from NRC Word-Emotion Association Lexicon \cite{mohammad2013nrc}. NRC Emotion Lexicon is a list of words and their associations with 8 basic emotions (anger, fear, anticipation, trust, surprise, sadness, joy, and disgust) and 2 sentiments (negative and positive). We classify users into the Big Five personality types which have been widely used to describe and predict users' activity on social media platforms \cite{souri2018personality}, \cite{utami2021personality}. We then adopt the continuous bag-of-words model commonly called word2vec to generate an empathy score for each user \cite{mikolov2013efficient}. Word2vec is a model that learns a vector representation for each word by using a neural network language model that can be trained on billions of words. We also use user comments to construct user features by going through the user comments and featuring them using the TfIdf + NMF combination. We use the TextWiser library to generate the comments features \cite{textwiser2021}. The last set of user features are the aggregated comment vectors over all the posts the user has interacted with and whether the user has interacted with any posts from a given subreddit (topic). Since the aim is to recommend content that helps the user regulate their emotions, we also took into consideration the popular subreddits the users have or have not subscribed to, to enable distraction, avoidance and relaxation as a part of emotion regulation. We then used \href{https://github.com/fidelity/selective?tab=readme-ov-file}{Selective}, a white-box feature selection library to help determine the subset of tabular user features that are most relevant for predicting users’ item preferences.

\subsubsection{Item}
The posts that the users interacted with were considered to be a set of items. The attributes associated with every post that was downloaded using the Reddit API are the post ID, its author, if it was clicked, comments, number of comments, list of authors in the comments, title, score, upvote ratio, subreddit etc. Apart from this, the items were mapped with the emotion that was expressed in their content as well as their emotional tone which is the set of prominent emotions expressed in their comments along with the statistics of each post which became the item features. We then go through the items (posts) dataset, calculating the most popular subreddits. We use the top 120 most common subreddits as topics. We create binary features for all posts based on whether a post contains one of the top 120 subreddits. We then look at the user comments and limit the user base to the subset where the users have both comments and interactions.  

\subsubsection{Interaction}
We generate the interaction data from users' upvotes and downvotes as responses on the platform. 
We categorised the interactions with binary labels where the response to an interaction is considered 1 if the user upvoted an item (post)  and 0 otherwise. Each interaction can be represented as a tuple ($u$, $i$, $r$) where $u$ is a user\_id, $i$ is an item\_id and $r$ is the observed response. We then perform cross-fold validation for training and testing. There were ~8000 items (posts) in the train and test sets with an average response rate of approximately 36\%. 

\subsection{Training}

The Contextual Multi-Armed Bandit (CMAB) is a reinforcement learning algorithm that uses contextual information to optimise rewards in dynamic environments. It excels in decision-making by observing contexts, choosing actions, and receiving specific rewards \cite{strehl2010learning}, \cite{mehrotra2020bandit}, \cite{lu2010contextual}, \cite{sutton2018reinforcement}. Contextual bandits are valuable in search and recommendation systems as they consider user-related data and address uncertainty by exploring unexplored items. These bandits adaptively focus on higher reward items and aim for personalised decision-making. Multi-Armed Bandits (MAB) balance exploration and exploitation in sequential decision-making, defining each arm as a decision an agent can make, with a deterministic or stochastic reward. 

In recommender systems, the arms correspond to different system items, and the reward is based on user interactions. Contextual Multi-Armed Bandits (CMAB) utilise contextual information affecting the reward for a given arm \cite{mcinerney2018explore}, \cite{he2020contextual}. Following describes the CMAB algorithm: 

At each time step \( t \), the environment presents a context \( x_t \) to the algorithm (often called the agent). Based on this context, the agent chooses an action \( a_t \) from a set of possible actions \( A \).

In response to the action, the agent receives a reward \( r_t \). The goal of the agent is to learn a policy \( \pi \), a mapping from contexts to actions, that maximises the cumulative reward over time:

\[
\max_{\pi} \mathbb{E} \left[ \sum_{t=1}^{T} r_t \mid \pi, D \right] \tag{1}
\]

where \( D = (x_1, a_1, r_1), \ldots, (x_T, a_T, r_T) \) is the data (context, action, reward) collected until time \( T \). The expectation is taken over the randomness in the context and rewards.


In this study, the arms represent different emotion regulation strategies (Empathic Responding, Distraction, Avoidance, Expression, and Relaxation). These strategies are tailored to individual user needs by considering user-specific features derived from their interaction history and current emotional state. The use of Contextual Multi-Armed Bandits (CMABs) is driven by the aim to utilise rich contextual information to enhance the relevance and effectiveness of recommendations \cite{schulz2015learning}, \cite{ban2021local}. Emotion regulation is a process heavily influenced by context, where the success of a strategy can vary significantly based on the user's current emotional state, previous interactions, and situational factors. CMABs adeptly handle these nuances, rendering them well-suited for our empathic content recommendation system.

In the context of emotion regulation (ER) recommendations, we formulate the problem as a CMAB using data from a content consumption application (Reddit). This includes users' demographic information, content preferences, user profile statistics, personality attributes, and previous posts,  that were processed into the contexts, the recommended ER strategies, and the associated rewards for our CMAB framework.

To assess the effectiveness of our policy, we rely on expected rewards obtained from testing data through offline learning. Offline evaluation replay is a technique used to evaluate the performance of a recommendation system by utilising historical data. In this method, the system generates recommendations using a specific algorithm and then compares them with historical data or ground truth \cite{tran2021combining}, \cite{ameko2020offline}. This technique is commonly used for evaluating contextual bandit algorithms.

For this study, we conducted an experiment to train the CMAB recommendation system utilising three prominent contextual bandit algorithms, subsequently comparing their performance \cite{kadioglu2024mab2rec}, \cite{strong2019mabwiser}.

\subsubsection{LinTS}The Linear Thompson Sampling (LinTS) algorithm is a Bayesian approach to contextual multi-armed bandits that is valuable in situations with high uncertainty and sparse data, as it effectively balances between exploring new options and exploiting known strategies. When it comes to emotional regulation, where user preferences and emotional responses can vary greatly, LinTS ensures that the system explores a wide range of decisions, enhancing its ability to identify effective strategies and avoiding repeatedly recommending the same items for similar emotional states.

This algorithm is based on Bayesian principles and extends Thompson Sampling to incorporate contextual information \cite{agrawal2013thompson}, \cite{kadioglu2024mab2rec}. It maintains a probabilistic model of the expected rewards for each option and updates this model based on observed rewards and contexts. The algorithm operates as follows: 

Initialisation: \begin{itemize}
    \item For each arm \(a\), initialise the mean vector \(\mu_a\) to zero.
    \item Initialise the covariance matrix \(\Sigma_a\) to a scaled identity matrix \(\lambda \mathbf{I}\), where \(\lambda > 0\) is the regularisation parameter and \(\mathbf{I}\) is the identity matrix.
\end{itemize}

Action Selection: At each time step \(t\):
\begin{itemize}
    \item For each arm \(a\), sample the parameter vector \(\theta_a\) from the multivariate normal distribution \(\mathcal{N}(\mu_a, \Sigma_a)\):
    \[
    \theta_a \sim \mathcal{N}(\mu_a, \Sigma_a)
    \]
    \item Compute the expected reward for each arm using the context vector \(x_t\):
    \[
    \hat{r}_a = x_t^T \theta_a
    \]
    \item Select the arm \(a_t\) with the highest expected reward:
    \[
    a_t = \arg\max_a \hat{r}_a
    \]
\end{itemize}

Model Update: After selecting arm \(a_t\) and observing the reward \(r_t\):
\begin{itemize}
    \item Update the covariance matrix \(\Sigma_a\) and the mean vector \(\mu_a\) for the selected arm \(a_t\) using Bayesian updating rules:
    \[
    \Sigma_a^{-1} \leftarrow \Sigma_a^{-1} + x_t x_t^T
    \]
    \[
    \mu_a \leftarrow \Sigma_a \left(\Sigma_a^{-1} \mu_a + r_t x_t\right)
    \]
\end{itemize}

\subsubsection{LinUCB}The Linear Upper Confidence Bound (LinUCB) algorithm is a highly effective approach for contextual multi-armed bandits, particularly valuable when the reward is presumed to be a linear function of the context vector. It uses linear regression (ridge regression) to estimate the reward function. LinUCB utilises confidence bounds to strike a balance between exploration and exploitation, making it well-suited for scenarios involving structured user feedback \cite{li2010contextual}, \cite{carpentier2011upper}. This algorithm works for our application in emotion regulation through digital content recommendation as it utilises contextual information to customise recommendations to individual users, ensuring that content aligns with their unique preferences and emotional needs. The algorithm efficiently handles binary rewards, making it ideal for our scenario where user feedback (e.g., likes or dislikes) is binary. LinUCB’s design allows it to effectively update its estimates and confidence bounds based on such feedback. LinUCB has the ability to adapt to evolving user behaviours and preferences by continuously updating its model with new data, ensuring that recommendations remain relevant and effective. By incorporating confidence bounds in its action selection process, LinUCB maintains equilibrium between exploring new content and exploiting known preferences, which is essential for discovering effective emotion regulation strategies \cite{yan2022dynamic}, \cite{kadioglu2024mab2rec}. The algorithm can be scaled to the large datasets typical of digital platforms and facilitating real-time recommendation updates based on user interactions. The algorithm operates as follows: 

Initialisation: \begin{itemize}
    \item For each arm \(a\), initialise the design matrix \(A_a = \mathbf{I}\), where \(\mathbf{I}\) is the identity matrix.
    \item Initialise the reward vector \(b_a = \mathbf{0}\), a zero vector.
\end{itemize}

Action Selection: At each time step \(t\):
\begin{itemize}
    \item For each arm \(a\), compute the parameter vector \(\theta_a\) using ridge regression:
    \[
    \theta_a = A_a^{-1} b_a
    \]
    \item Calculate the upper confidence bound for each arm:
    \[
    \hat{r}_a = x_t^T \theta_a + \alpha \sqrt{x_t^T A_a^{-1} x_t}
    \]
    where \(x_t\) is the context vector and \(\alpha\) is a parameter that controls the degree of exploration.
    \item Select the arm \(a_t\) with the highest upper confidence bound:
    \[
    a_t = \arg\max_a \hat{r}_a
    \]
\end{itemize}

Model Update: After selecting arm \(a_t\) and observing the binary reward \(r_t\):
\begin{itemize}
    \item Update the design matrix \(A_a\) and the reward vector \(b_a\) for the selected arm \(a_t\):
    \[
    A_a \leftarrow A_a + x_t x_t^T
    \]
    \[
    b_a \leftarrow b_a + r_t x_t
    \]
\end{itemize}

\subsubsection{LogUCB}
The LogUCB (Logistic Upper Confidence Bound) algorithm is specifically designed to handle situations where rewards are binary and modeled using logistic regression. This algorithm effectively combines the principles of upper confidence bounds (UCB) and logistic regression to strike a balance between exploration and exploitation. Our current scenario involves binary interactions, such as upvote/downvote, (where user engagement is assessed in binary term), as mentioned in part 3 of Data Collection. This makes logistic regression a suitable model for estimating the probability of a positive response. LogUCB is an extension of the UCB algorithm, tailored to handle binary reward data using a logistic regression model \cite{mahajan2012logucb}, \cite{carpentier2011upper}. By integrating upper confidence bounds, LogUCB effectively manages the trade-off between exploration and exploitation, ensuring continual learning about less-explored areas while optimising for user engagement. The incorporation of context vectors allows the algorithm to take into account user-specific and content-specific features, ultimately leading to more personalised recommendations \cite{yan2022dynamic}, \cite{kadioglu2024mab2rec}. The regularised logistic regression approach employed by LogUCB is not only computationally efficient but also scalable, making it well-suited for handling large datasets in digital platforms. This algorithm operates as follows: 

Initialisation: \begin{itemize}
    \item Initialise the parameter vector \(\theta_0\) to zero.
    \item Set the regularisation parameter \(\lambda > 0\) to ensure numerical stability.
    \item Initialise the covariance matrix \(\mathbf{A} = \lambda \mathbf{I}\), where \(\mathbf{I}\) is the identity matrix.
    \item Initialise the gradient vector \(\mathbf{b} = \mathbf{0}\).
\end{itemize}

Action Selection: At each time step \(t\):
\begin{itemize}
    \item Compute the posterior distribution of the parameters using the current covariance matrix and gradient vector: \(\mathbf{A}^{-1} = (\mathbf{A})^{-1}\) and \(\hat{\theta}_t = \mathbf{A}^{-1} \mathbf{b}\).
    \item For each arm \(a\), calculate the upper confidence bound (UCB) for the logistic model:
    \[
    \text{UCB}_a(t) = \hat{x}_a^T \hat{\theta}_t + \alpha \sqrt{\hat{x}_a^T \mathbf{A}^{-1} \hat{x}_a}
    \]
    where \(\hat{x}_a\) is the context vector for arm \(a\) at time \(t\) and \(\alpha\) is a parameter that controls the trade-off between exploration and exploitation.
    \item Select the arm \(a_t\) with the highest UCB value:
    \[
    a_t = \arg\max_a \text{UCB}_a(t)
    \]
\end{itemize}

Model Update: After selecting arm \(a_t\) and observing the reward \(r_t \in \{0, 1\}\):
\begin{itemize}
    \item Update the covariance matrix \(\mathbf{A}\) and gradient vector \(\mathbf{b}\):
    \[
    \mathbf{A} \leftarrow \mathbf{A} + \hat{x}_{a_t} \hat{x}_{a_t}^T
    \]
    \[
    \mathbf{b} \leftarrow \mathbf{b} + r_t \hat{x}_{a_t}
    \]
    \item Update the parameter vector \(\hat{\theta}_t\) using the new \(\mathbf{A}\) and \(\mathbf{b}\):
    \[
    \hat{\theta}_t \leftarrow \mathbf{A}^{-1} \mathbf{b}
    \]
\end{itemize}

\begin{table*}[]
\centering
\caption{Survey questions for the user study}
\label{tab:survey_questions}
\resizebox{\textwidth}{!}{%
\begin{tabular}{ll}
\hline
\textbf{Question} & \textbf{Responses} \\ \hline
Emotional State & Angry, fearful, happy, joyful, sad, surprise, neutral, other \\ 
Intensity of Current Emotion & A little, moderate amount, a lot, can't say \\ 
Modify Current Emotional State? & No, Yes, Maybe \\ 
Can Interaction with Others Help? & No, Yes, Maybe \\ 
Interest in Digital Platform Influence? & No, Yes, Maybe \\ 
Preferred Interaction on Social Media & Content descriptions based on emotion regulation strategies \\ 
Reason for Engagement & Various reasons, including expression, emotional tone, timeliness, relatable themes, etc. \\ 
Avoid Interacting with Specific Content & Content descriptions based on emotion regulation strategies \\
Reason for Avoidance & Various reasons, including inappropriate tone, timing, lack of connection, irrelevant themes, etc. \\ \hline
\end{tabular}%
}
\end{table*}

\subsection{Evaluation}
We compared a set of recommender algorithms by training, scoring, and evaluating each algorithm using cross-validation using offline learning. To evaluate the recommendation algorithms, we used the generated recommendations on historical data and predicted the expected reward for those recommendations. We also utilised widely accepted metrics such as Area Under the Curve (AUC), Click-Through Rate (CTR), Precision, and Recall to evaluate the performance of our recommendation systems. These metrics are well-documented in the literature as standard measures for assessing the effectiveness of recommendation systems. \cite{zeng2016online}, \cite{ayata2018emotion}, \cite{saraswat2020analyzing}.

\subsection{User survey}
We conducted a user study to assess how users responded to the empathetic content generated by the CMAB recommender. To ensure accuracy and reliability, we utilised a survey method that required participants to complete a brief questionnaire three times per day (at 10 AM, 1 PM and 5 PM) for seven consecutive days. Surveys have been widely used to evaluate recommendation systems as they provide a comprehensive and user-centric approach to evaluating recommendation systems \cite{tag2021retrospective}. We followed a rigorous ethics approval (at Deakin University, Australia) process to ensure the well-being of our participants, as the data collected was personal in nature. The study aimed to focus on Australian adults who frequently use social media, as they are both an important demographic and convenient for sampling purposes. We recruited 20 participants from different social media platforms, out of which 14 were either employed full-time or part-time, and the remaining 6 were students. Our chosen seven-day duration allowed us to capture a broad range of experiences while still ensuring precise responses. 

Table. \ref{tab:survey_questions} outlines the survey questions that were created to gain insights into users' emotional states, their preferences for emotional regulation strategies, and their interactions with digital platforms. Participants were initially prompted to select their current emotional state from a list of six primary emotions that included anger, fear, happiness, joy, sadness, surprise as well as neutral, or other. After this, they were asked to rate the intensity of their emotions on a scale from ``a little" to ``a lot" or choose ``can't say." The survey went on to explore whether participants were interested in modifying their emotional state or preferred to stay as is or were unsure. It also examined participants' views on the role of interpersonal interactions in managing emotions, determining whether engaging with others could be helpful. Participants were asked to specify which types of content descriptions they preferred based on their emotion regulation strategies and to explain their reasons for engaging with particular content. Additionally, if participants avoided certain content descriptions, they were asked to explain their reasons, which could range from inappropriate tone and timing to lack of connection and irrelevant themes. These questions aimed to provide an understanding of users' emotional experiences, preferences, and digital engagement behaviours about emotional regulation.

As a token of our appreciation, we offered a \$50 Coles (a popular Australian supermarket chain) voucher to each participant who completed the study. We then analysed the data to gain insights into the effectiveness of the system in delivering empathetic content and positively impacting users' emotional experiences. The results of our analysis offer valuable insights into the system's ability to provide empathetic content that resonates with users.

  


\section{Results and Discussion}

The results and discussion section provides an analysis of the assessment of the Contextual Multi-Armed Bandit (CMAB) recommendation system, as well as insights obtained from the user survey. This section aims to describe the effectiveness of the proposed framework in regulating emotions through empathetic content recommendations and to comprehend user preferences and responses to the content generated for system-initiated interpersonal ER.
\begin{figure*}[htbp]
  
    \centering
    \includegraphics[width=15cm,height=12cm,keepaspectratio]{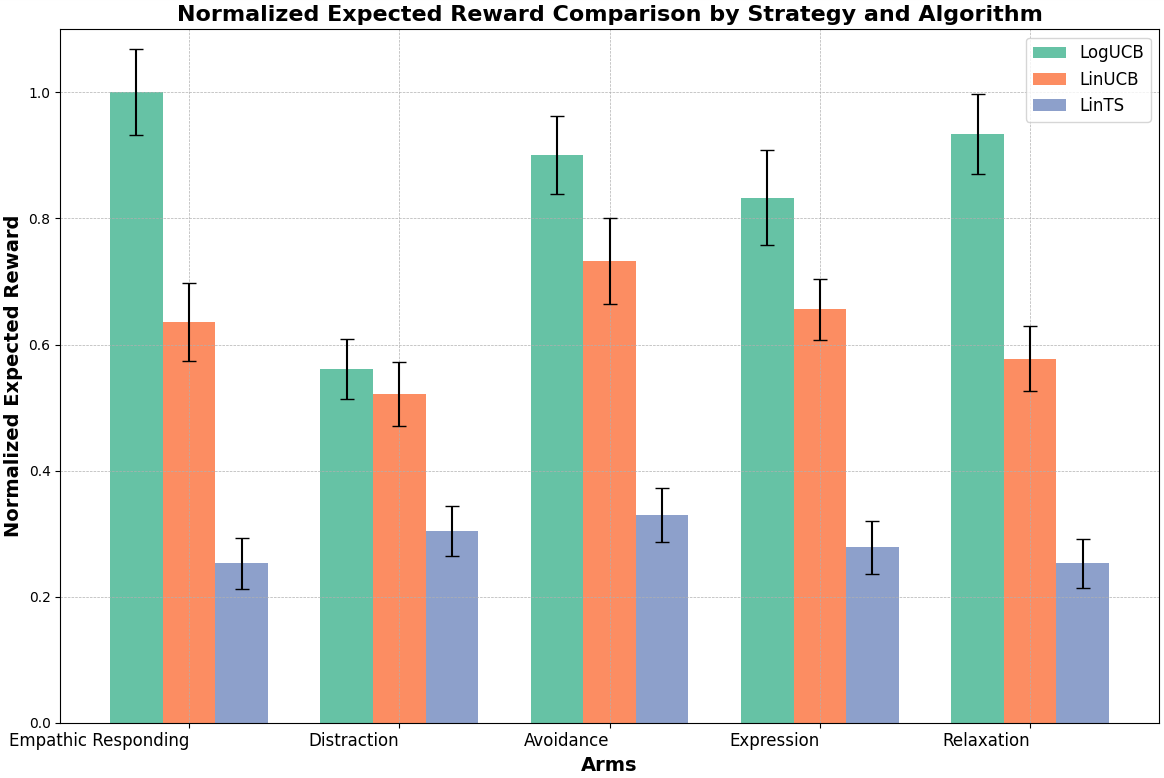}
  \caption{Average expected rewards for each ER strategy and policy}
  \label{fig:exp_rewards}
  \end{figure*} 

\subsection{CMAB Recommendation System}
We evaluated the optimal strategy for generating empathic recommendations by analysing the expected rewards across 100 recommendations per learning policy. The recommended emotion regulation (ER) strategy for each instance was identified by using the keys from these recommendations. The expected rewards were then sorted based on the ER strategy (arms) and plotted against the average rewards for each strategy and policy, namely Empathic Responding, Distraction, Avoidance, Expression, and Relaxation. The results revealed that the LogUCB policy outperformed the other evaluated policies for four out of five arms of the Contextual Multi-Armed Bandit (CMAB).

The performance hierarchy among the tested policies was distinct, as illustrated in the corresponding plot Fig. \ref{fig:exp_rewards} where the x-axis denotes the five arms of the CMAB and the y-axis denotes the expected rewards across 100 recommendations per learning policy. The LogUCB policy achieved the highest average expected rewards, outperforming the other strategies across four arms while LinUCB performed best for Expression. The LogUCB method applies a logarithmic transformation, enabling it to effectively capture non-linear relationships between features and binary rewards. This capability is particularly beneficial for our feature set, which encompasses user behaviours, emotional states, and interaction patterns. By accommodating non-linearities, LogUCB can make more accurate predictions. In contrast, both LinUCB and LinTS rely on the assumption of a linear relationship between context (features) and rewards, which may not fully capture the complex relationships within our data, potentially resulting in less accurate recommendations \cite{mahajan2012logucb}, \cite{yan2022dynamic}, \cite{agrawal2013thompson}.

As our feature engineering process involved extracting user attributes, such as inferred emotions, personality types, and engagement with specific subreddits. These features likely introduce complex, non-linear interactions, which LogUCB is well-equipped to leverage. For example, the interaction between emotional tone and personality type can impact a user's response to different emotion regulation strategies. The flexibility of LogUCB in handling such interactions likely contributes to its better performance.

  

\begin{figure*}[ht]
    \centering
    \subfloat[\label{fig:image1}]{%
        \includegraphics[width=0.45\textwidth, height=8cm]{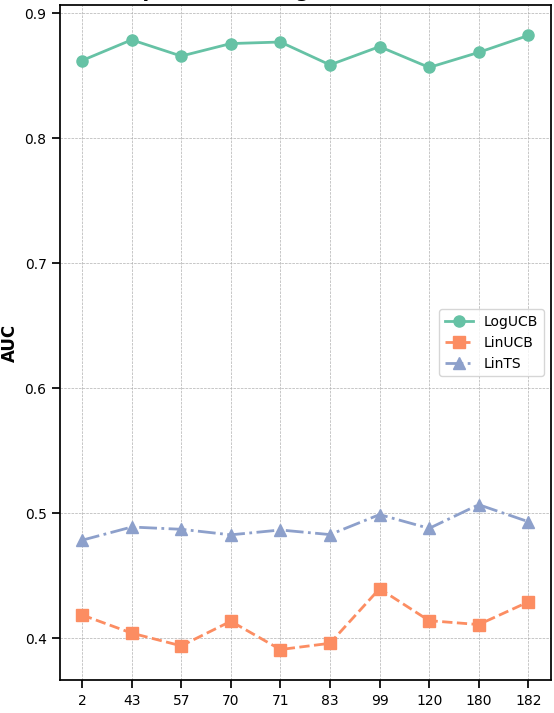}}
    \subfloat[\label{fig:image2}]{%
        \includegraphics[width=0.45\textwidth, height=8cm]{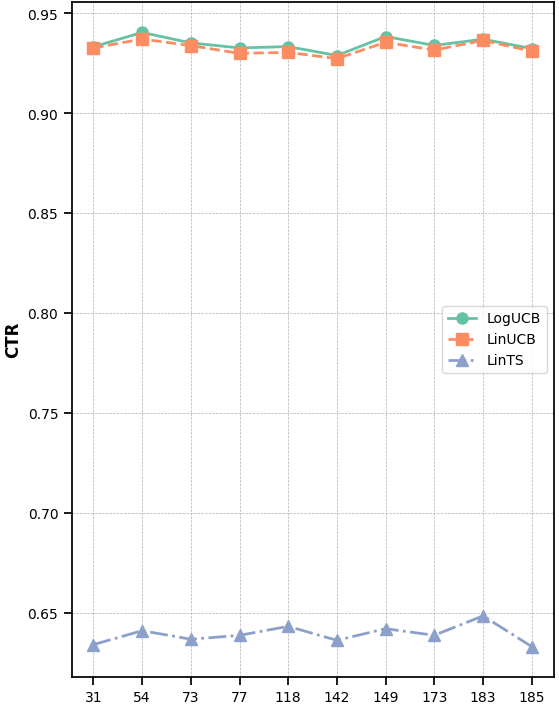}}\\
    \subfloat[\label{fig:image3}]{%
        \includegraphics[width=0.45\textwidth, height=8cm]{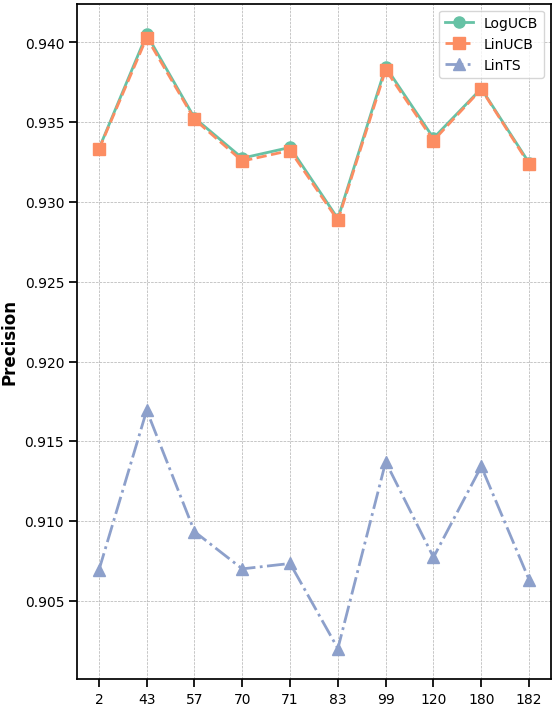} }
    \subfloat[\label{fig:image4}]{%
        \includegraphics[width=0.45\textwidth, height=8cm]{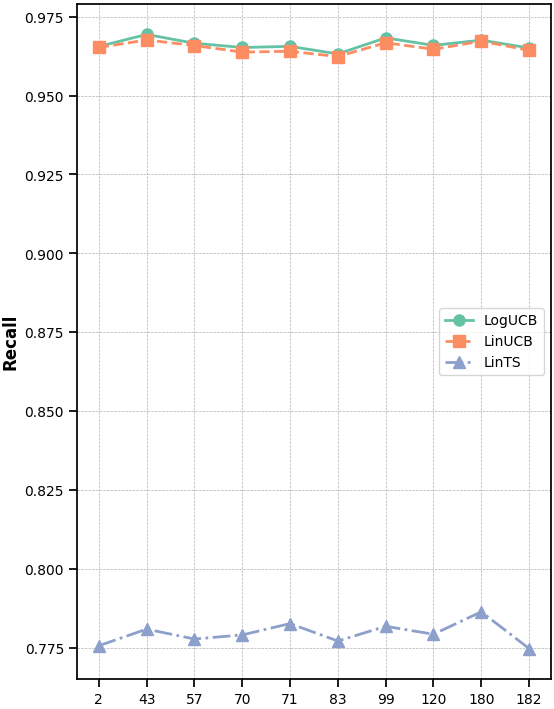}}
    
    \caption{Binary Recommender and Ranking metrics for each policy. AUC, CTR, Precision and Recall respectively.}
    \label{fig:metrics}
\end{figure*}
LogUCB inherently strikes a balance between exploration and exploitation by dynamically adjusting confidence bounds using a logarithmic function. This ensures effective exploration without over-exploring, ultimately leading to more stable and optimised recommendations over time.

While LinUCB also achieves a balance between exploration and exploitation, its linear approach may not be as effective in adapting to the varied contexts in our dataset. Similarly, LinTS which relies on sampling from a posterior distribution, may not explore as efficiently in our specific context, potentially resulting in suboptimal recommendations.

Since our dataset comprises both structured tabular data and unstructured text data. LogUCB's ability to handle a wide range of feature types and distributions likely contributes to its robustness and superior performance. In contrast, LinUCB and LinTS may struggle with the variability and noise present in the unstructured text features, such as comments and emotional content, leading to less consistent performance.

To provide a comprehensive assessment of the policies, binary recommender metrics (such as the Area under curve - AUC and Click-through rate - CTR) and ranking metrics (such as Recall and Precision) were employed, as demonstrated in Fig. \ref{fig:metrics}. The x-axis of Fig. \ref{fig:metrics} represents the 10 random top recommendations, while the y-axis displays the metric value for the corresponding top n recommendations. AUC is a commonly used metric in recommendation systems that calculates the probability of a randomly selected relevant/clicked item being ranked higher by the recommender than a randomly chosen non-relevant/not-clicked item. This calculation is performed over the subset of user-item pairs that appear in both actual ratings and recommendations. AUC is particularly useful in evaluating the performance of a recommendation system because it considers the ranking of all items, not just those that were recommended. The Click-Through Rate (CTR) metric is widely used to evaluate the effectiveness and accuracy of recommendation systems. It measures the proportion of user interactions with recommended items over the subset of user-item pairs that appear in both actual ratings and recommendations. This subset is also known as the ``intersection set," which serves as a benchmark for evaluating the quality of recommendations.

Precision refers to the model's ability to consistently identify items that a user is likely to interact with. For instance, if a recommendation system only provides recommendations for a small percentage of users, it can still achieve high precision if those recommendations consistently result in user interaction. On the other hand, recall measures the model's ability to capture all the items that a user has interacted with. Even if a recommendation system generates many irrelevant recommendations, it can still achieve high recall if it also identifies the relevant recommendations.

We found that, LogUCB consistently demonstrates strong performance across all the above metrics. It achieves an average precision of approximately 0.936, indicating its ability to accurately recommend relevant items. In terms of recall, LogUCB also performs well, with an average score of 0.966, showing its ability to capture relevant recommendations. The area under the curve (AUC) for LogUCB averages around 0.867, confirming its effectiveness in ranking recommendations. Additionally, LogUCB shows a high click-through rate (CTR) of approximately 0.937, highlighting strong user engagement compared to the other policies.

While LinUCB maintains competitive precision and recall scores, averaging around 0.933 and 0.965 respectively, its AUC and CTR metrics are notably lower. With an average AUC of about 0.421, LinUCB indicates a less effective ranking compared to LogUCB. Similarly, its CTR averages around 0.406, suggesting lower user interaction rates despite comparable precision and recall scores.

On the other hand, LinTS exhibits lower precision and recall metrics, averaging around 0.624 and 0.911 respectively. Its AUC is notably lower at approximately 0.476, indicating weaker performance in ranking relevant recommendations. The CTR for LinTS averages around 0.935, indicating a strong level of user engagement. 
Based on our evaluation, we conclude that LogUCB is the best performer among the three policies for our recommendation system and implement it for generating recommendations for user surveys.
\begin{figure*}[!htbp]
  
    \centering
    \includegraphics[width=18cm,height=15cm,keepaspectratio]{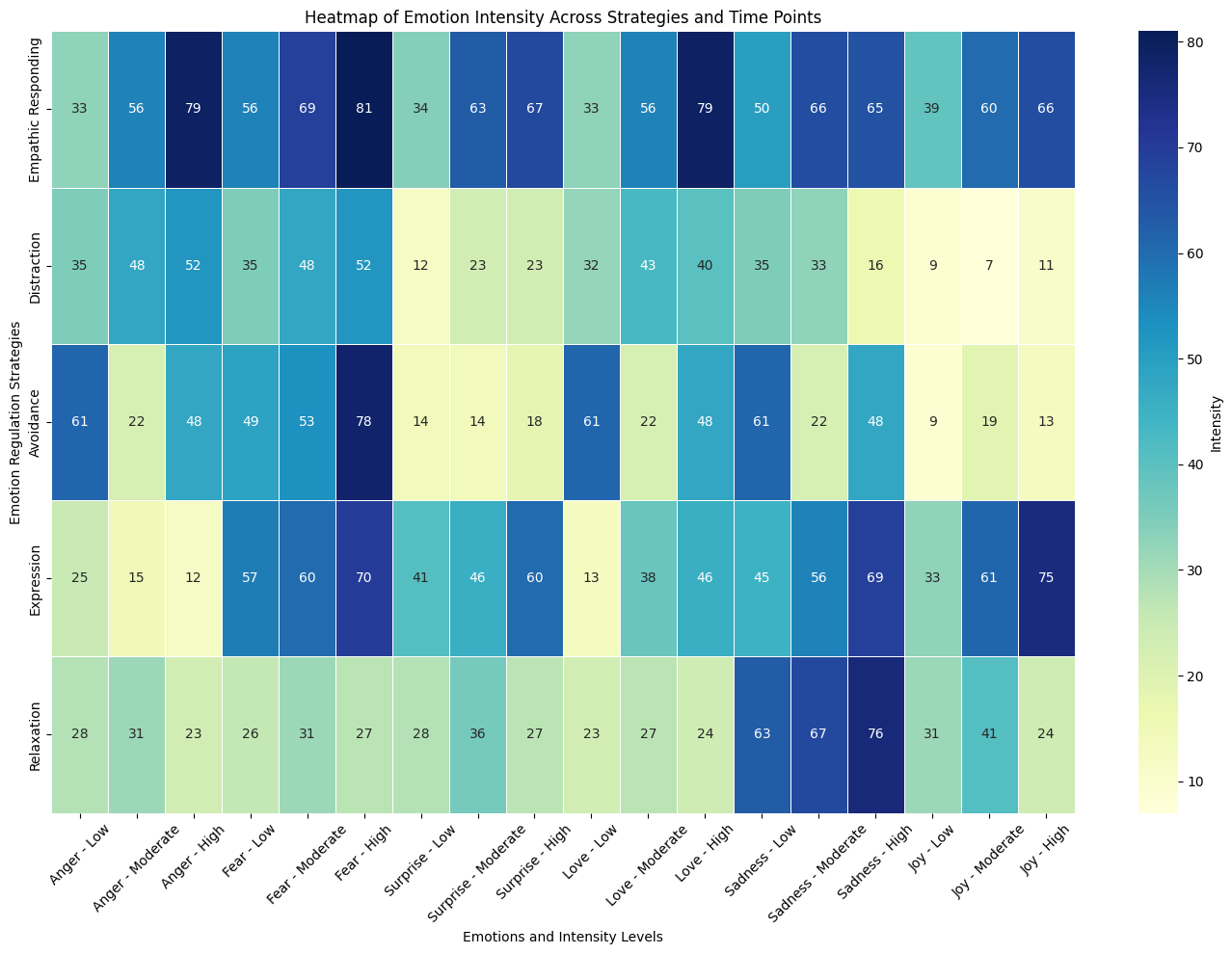}
  \caption{Heatmap of ER strategy preferences w.r.t emotions and their intensities}
  \label{fig:user_survey}
  \end{figure*}  


\subsection{User Survey}    
The effective management of emotional experiences is essential in various contexts, and emotion regulation strategies play a significant role in this process. To shed light on the effectiveness of empathic responding through content recommendations and contextual preferences for different emotion regulation strategies, our study surveyed respondents experiencing the six primary emotions, including anger, fear, surprise, love, sadness, and joy. Our findings reveal multifaceted patterns in the preference for specific strategies based on the type of emotion and its intensity.

We employed a heatmap to visually represent the strength of responses for different emotion regulation strategies, categorised by three levels of intensity (Low, Moderate, High) for each emotion (Anger, Fear, Surprise, Love, Sadness, and Joy). Fig. \ref{fig:user_survey} represents distinct patterns of emotional response across various strategies and emotions by displaying a range of response intensities. Based on our statistical analysis of user survey results, we've discovered a significant correlation between a user's current emotional state and their preferred emotion regulation strategies. It's important to note that these relationships aren't absolute. For example, respondents experiencing anger tended to prefer empathic responding, avoidance, and distraction, while those experiencing joy or surprise leaned towards expression and relaxation strategies. We also found a connection between the intensity of emotions being experienced and the regulation strategy chosen. For instance, 78\% of respondents experiencing moderate or high-intensity emotions preferred empathic responding, compared to 43\% of those with low-intensity emotions. Interestingly, 96\% of respondents were willing to regulate their emotions and open to digital content recommendations, even if they didn't feel the need to regulate at the given moment. Additionally, 100\% of respondents who believed interacting with someone could help improve their emotional state were open to engaging in emotion regulation via digital content recommendations. 

As can be observed from Fig. \ref{fig:user_survey}, strategies such as ``Empathic Responding" and ``Relaxation" consistently demonstrate higher intensity levels across most emotions, particularly in the Moderate to High categories. This suggests that these strategies may have broader applicability or effectiveness in regulating emotional states. On the other hand, strategies like ``Distraction" and ``Avoidance" exhibit varying degrees of effectiveness, with some emotions such as Sadness and Anger demonstrating lower intensities in the Low category, indicating possible limited utility depending on the emotional context. Additionally, specific emotions such as Sadness and Love display notably high intensities under the ``Expression" strategy, indicating a potential inclination or effectiveness in expressing these emotions as a coping mechanism. Conversely, strategies like ``Avoidance" exhibit an interesting distribution with high intensities in some emotions (e.g., Fear) but lower intensities in others (e.g., Joy), suggesting that the appropriateness of avoidance varies significantly depending on the nature of the emotion. 

Therefore, the study revealed that empathic responding was a preferred strategy for managing four of the six emotions, with participants showing a tendency to engage in empathetic responses, especially during moments of heightened emotional intensity. These results emphasise the significant role of social connections and support in emotional regulation across a wide range of emotional experiences. Meanwhile, distraction, which involves redirecting one's focus from the source of emotional arousal, consistently showed moderate preference levels across various emotions. However, its overall effectiveness appeared to be slightly lower than other techniques, indicating that its usefulness may depend on the intensity of the emotion being experienced. We also found that respondents were less inclined to use avoidance as a primary emotion regulation strategy, as it was the least preferred approach for all emotions. This suggests that avoiding emotional distress may not be an effective long-term coping mechanism. On the other hand, expression, whether through videos or text, was a preferred strategy for respondents experiencing a range of emotions including surprise, love, sadness, and joy. This indicates that individuals tend to seek opportunities to express their emotions as a way of processing and regulating their emotional experiences. Furthermore, relaxation techniques aimed at reducing stress and promoting relaxation were consistently preferred across different emotions. These findings highlight the importance of conscious practices in managing emotional well-being, regardless of the specific emotional context.
The study provides valuable insights into the utilisation of different emotion regulation strategies across various emotional experiences. These insights contribute to our understanding of adaptive emotion regulation processes and have implications for the development of targeted interventions aimed at enhancing emotional well-being and resilience.

\section{Limitations}
The present study provides insights into the potential of a proposed recommendation system for facilitating effective emotion regulation on digital platforms. However, several limitations warrant consideration. Firstly, the study focused on evaluating a subset of emotion regulation strategies, which may not provide a complete understanding of emotion regulation dynamics in online environments. Future research could explore the integration of additional strategies to enhance the robustness and applicability of the proposed approach.

Secondly, the user study conducted to assess the efficacy of the recommendation system involved a relatively small sample size, which may constrain the generalization of the results. Future studies with larger and more diverse participant groups could provide further validation of the approach and its effectiveness across different user demographics and contexts.

Additionally, the study primarily relied on text-based social media data from Reddit to design and evaluate the recommendation system. While Reddit serves as a rich source of user-generated content and interactions, the findings may not fully capture the dynamics of emotion regulation across various digital platforms and communication mediums. Future research could extend the scope of the study to include multiple platforms and data types to enhance the robustness and applicability of the proposed approach.

Lastly, the proposed recommendation system is designed to operate within the constraints and limitations of existing digital media platforms. The effectiveness of the system may be influenced by external factors such as platform policies, algorithms, and user interface designs. Further refinement and adaptation of the system to address these external variables could enhance its overall performance and user acceptance.

Overall, while the current study presents a step towards enhancing Interpersonal Emotion Regulation (IER) on digital platforms, the aforementioned limitations highlight the need for continued research and development to optimise the effectiveness and applicability of the proposed approach in real-world settings.

\section{Conclusion}
The ever-changing landscape of digital communication has made online platforms increasingly important in facilitating interpersonal emotion regulation (IER). This study aimed to address the gap in existing digital platforms, which often fall short of supporting effective emotion regulation. We introduced a novel recommendation system designed to promote empathic responding strategies and enhance IER on digital platforms.
The proposed recommendation system demonstrated promising results through the analysis of 37.5K user posts and interactions from Reddit. Surveyed users consistently preferred empathic recommendations over other emotion regulation strategies such as distraction and avoidance. This preference highlights the importance of tailored content that aligns with users' emotional needs and preferences.

Our approach addresses users' emotional needs and preferences, effectively bridging the gap between system-initiated and user-initiated emotion regulation. It provides a real-time solution for fostering effective IER practices on digital media platforms. By incorporating both user content preferences and emotional alignment, our recommendation system enhances empathic and personalised online interactions. This research lays the groundwork for emotion regulation in digital media, guiding the development of more supportive online environments. As digital applications increasingly shape our online experiences, integrating effective emotion regulation strategies will be essential for promoting positive interactions and improving user well-being.

\section*{}
During the drafting of this paper, Grammarly \cite{grammarly} was used to check and enhance the grammar and writing style of this document.
\bibliographystyle{IEEEtran}
\bibliography{refs.bib}

\vfill

\end{document}